# 648 Hilbert space dimensionality in a biphoton frequency comb

Kai-Chi Chang[1,+,*], Xiang Cheng[1,2,+], Murat Can Sarihan[1], Abhinav Kumar[1], Yoo Seung Lee[1], Tian Zhong[3], Yan-Xiao Gong[4], Zhenda Xie[5], Jeffrey H. Shapiro[6], Franco N. C. Wong[6], and Chee Wei Wong[1,*]

[1] Fang Lu Mesoscopic Optics and Quantum Electronics Laboratory, Department of Electrical Engineering, University of California, Los Angeles, CA 90095, USA

[2] State Key Laboratory of Information Photonics and Optical Communications, Beijing University of Posts and Telecommunications, Beijing 100876, PR China

[3] Institute for Molecular Engineering, University of Chicago, Chicago, Illinois 60637, USA

[4] National Laboratory of Solid State Microstructures and School of Physics, Nanjing University, Nanjing 210093, PR China

[5] National Laboratory of Solid State Microstructures and School of Electronic Science and Engineering, Nanjing University, Nanjing 210093, PR China

[6] Research Laboratory of Electronics, Massachusetts Institute of Technology, Cambridge, MA 02139, USA

+ equal contribution

* email: uclakcchang@ucla.edu; cheewei.wong@ucla.edu

**Qubit entanglement is a valuable resource for quantum information processing, where increasing its dimensionality provides a pathway towards higher capacity and increased error resilience in quantum communications, cluster computation and quantum phase measurements. Time-frequency entanglement, a continuous variable subspace, enables the high-dimensional encoding of multiple qubits per particle, bounded only by the spectral correlation bandwidth and readout timing jitter. Extending from a dimensionality of two in discrete polarization variables, here we demonstrate a hyperentangled, mode-locked, biphoton frequency comb with a time-frequency Hilbert space dimensionality of at least 648. Hong-Ou-Mandel revivals of the biphoton qubits are observed with 61 time-bin recurrences, biphoton joint spectral correlations over 19 frequency-bins, and an overall interference visibility of the high-dimensional qubits up to 98.4%. We describe the Schmidt mode decomposition analysis of the high-dimensional entanglement, in both time- and frequency-bin subspaces, not only verifying the entanglement dimensionality but also examining the**

**time-frequency scaling. We observe a Bell violation of the high-dimensional qubits up to 18.5 standard deviations, with recurrent correlation-fringe Clauser-Horne-Shimony-Holt S-parameter up to 2.771. Our biphoton frequency comb serves as a platform for dense quantum information processing and high-dimensional quantum key distribution.**

Qubit entanglement is a fundamental differentiator of quantum information processing over classical systems – in computing, communications, simulations, and metrology. The dimensionality and information capacity encoded in entanglement can be scaled by the number of physical-logical qubits such as in multi-partite systems [1-11], and the dimensionality of each qubit. Entanglement dimensionality [12-19] can be addressed in spatio-temporal-energy properties such as orbital angular momentum [20, 21], position-momentum [22-25], polarization [26, 27], energy-time [28-30] and time-frequency [31-34] including for *d*-dimensional cluster computation [11, 34]. In quantum communications such as device-independent quantum cryptography, high-dimensionality provides a pathway towards increased information capacity per photon [12, 19, 30, 35-38], improved security against different attacks [39], and better resilience against noise and error [40, 41]. Consequently, advances in high-dimensional encoding of qubits have ranged from Bell-type inequalities for energy-time qudits [20, 34, 42, 43], to on-chip quantum frequency comb generation [32, 44], certifying high-dimensional entanglement via two global product bases without requiring large full state tomography [45], and increased noise resilience with larger noise fractions [40, 41]. Dimensionality of such robust qubits involved, for example, compressive sensing of joint quantum systems with 65,536 dimensions in the position-momentum degree-of-freedom [22], on-chip frequency-bin generation with at least 100 dimensions [32], and 19 time-bin qubits in biphoton frequency combs [31], scaling remarkably to even 84th time-bin revivals with a path length difference of 100 m [46]. To date, however, the Hilbert space dimensionality of time-frequency entanglement has always been less than 100 and the Schmidt mode decomposition of such dual time-frequency qubits has not yet been fully explored.

Here we demonstrate and quantify a hyperentangled biphoton frequency comb (BFC) with a time-frequency dimensionality of at least 648. First, we observe periodic time-bin revivals of the high-dimensional Hong-Ou-Mandel indistinguishability interference, scaling over 61 time-bins and with visibility up to 98.4%. Second, with time-frequency duality in the quantum frequency comb, we measure the joint spectral correlations over 19 frequency-bins, witnessing the spectral-temporal high-dimensional entanglement generality across three different cavities that span nearly an order-of-magnitude in mode spacings. Third, with the qubit polarization subspace, we obtain a mean visibility of the entanglement correlation fringes up to 2.771 ± 0.016, along with Bell violation of the high-dimensional qubits up to 18.5 standard deviations. Fourth, we perform the Schmidt mode decomposition analysis of the high-dimensional time-frequency qubits, through the dual and discretized joint spectral-temporal intensities. In the frequency subspace, the Schmidt-mode eigenvalues are described and extracted over three different cavities to quantify their effective dimensions, as well as to estimate their bounds. In the temporal subspace, we arrived at the time-bin Schmidt-mode eigenvalues, along with theory-experimental dimensionality quantification and their scaling. Including the polarization subspace, we achieved a Hilbert space dimensionality of at least 648 – to the best of our knowledge, this represents the highest Schmidt number value observed so far in time-frequency entanglement. Our results support the hyperentangled high-dimensional BFC as a platform for hybrid time-frequency quantum key distribution, time-frequency cluster state computations, and dense quantum information processing.

**Results**

**Time-bin revival subspaces:** Figure 1a shows the experimental setup. Spontaneous parametric downconversion (SPDC) occurs in a type-II periodically-poled KTiOPO$_4$ (ppKTP) waveguide, integrated in a fiber package for high fluence and efficiency [47]. It is pumped by a 658 nm wavelength Fabry-Pérot laser diode, stabilized by self-injection-locking. The type-II quasi-phase-matching is designed for generating orthogonally-polarized frequency-degenerate signal and idler photon pairs at 1316 nm with 245 GHz full-width-to-half-maximum (FWHM) phase-matching bandwidth [48]. The high-dimensional mode-locked two-photon state [31] is created by passing the signal and idler photons through one of three Fabry-Pérot fiber cavities, whose FSRs are 45.32 GHz, 15.15 GHz, and 5.03 GHz, with FWHM linewidths of 1.56 GHz, 1.36 GHz, and 0.46 GHz respectively. Each fiber cavity is mounted on a modified thermoelectric assembly with ≈ 1 mK temperature-control stability. A frequency-stabilized tunable reference laser at 1316 nm is used to align each cavity's spectrum to the SPDC's degenerate frequency by matching its second-harmonic to the 658 nm pump wavelength. The BFC state generated in this manner can be expressed as:

$$|\psi\rangle = \sum_{m=-N}^{N} \int d\Omega\, f(\Omega - m\Delta\Omega)\hat{a}_H^\dagger\left(\frac{\omega_p}{2} + \Omega\right)\hat{a}_V^\dagger\left(\frac{\omega_p}{2} - \Omega\right)|0\rangle, \qquad (1)$$

where $\Delta\Omega$ is the cavity's FSR in rad s$^{-1}$, $\Omega$ is the detuning of the SPDC biphotons from frequency degeneracy, $2N+1$ is the number of cavity lines passed by an overall bandwidth-limiting filter, and the state's spectral amplitude $f(\Omega - m\Delta\Omega)$ is the single frequency-bin profile defined by the cavity Lorentzian transmission lineshape with FWHM linewidth $2\Delta\omega$

$$f(\Omega) = \frac{1}{[(\Delta\omega)^2 + \Omega^2]}. \qquad (2)$$

The signal and idler photons are cleanly separated by a polarizing beamsplitter (PBS) by our type-II SPDC configuration, so that the BFC is generated without post-selection. Using the temporal wavefunction, the BFC's state can be rewritten as:

$$|\Psi\rangle = \int d\tau \exp(-\Delta\omega|\tau|) \frac{\sin[(2N+1)\Delta\Omega\tau/2]}{\sin(\Delta\Omega\tau/2)} \hat{a}_H^\dagger(t)\hat{a}_V^\dagger(t+\tau)|0\rangle. \qquad (3)$$

where the exponential decay is slowly varying relative to the $\sin[(2N+1)\Delta\Omega\tau/2]/\sin(\Delta\Omega\tau/2)$ term because $\Delta\omega \ll \Delta\Omega$. Hence, the BFC's temporal wavefunction has many peaks, with repetition period equal to the cavity round-trip time, $\Delta T = 2\pi/\Delta\Omega$, where $\Delta T \approx 22.1$ ps, 66.0 ps, and 198.8 ps for the BFCs generated by the 45.32 GHz, 15.15 GHz, and 5.03 GHz cavities respectively.

A fiber Bragg grating (FBG) of 346 GHz bandwidth with a circulator and a long-pass filter are used to spectrally select – setting the $N$ value in Eq. (1) – the high-dimensional BFC while blocking residual pump light. The BFC wavefunction in Eq. (3) implies that the HOM interference will have revivals at relative delays corresponding to integer multiples of the fiber cavity's round-trip time [31, 49, 50], which we experimentally verified as follows. The orthogonally-polarized signal and idler photons are separated by a PBS and directed to different arms of the HOM interferometer. A fiber polarization controller (FPC) in the lower arm of the interferometer rotates the idler photon's polarization to match that of the signal photon at the 50:50 fiber coupler. An optical delay line with less than 0.02 dB insertion loss over its 220 mm travel range is used to vary the relative delay between the signal and idler photons for quantum interference. After the HOM interferometer, coincidences are measured with two superconducting nanowire single photon detectors (SNSPDs, $\approx 85\%$ detection efficiency).

The HOM experimental results in Figure 1b are obtained with the 45.32 GHz FSR fiber cavity by scanning the relative optical delay between the two arms of the HOM interferometer from - 340 ps to +340 ps with respect to the central dip. A pump power of 2 mW is chosen to avoid the multi-pair emissions that decrease two-photon interference visibility [31, 51]. The fringe visibility of the quantum interference, $V_n$, for the $n$th dip is $[C_{\max} - C_{\min}(n)]/C_{\max}$, where $C_{\max}$ is the maximum coincidence count and $C_{\min}(n)$ is the minimum coincidence count of the *n*th dip. In the left inset of Figure 1b, the central bin visibility is observed to be 98.4% before subtracting accidental coincidences, and 99.9% after they are subtracted, supporting the good fidelity of our entangled state. The base-to-base width of the central dip is fitted to be 3.86 ± 0.30 ps, which agrees well with our 245 GHz phase-matching bandwidth. We obtained HOM-dip revivals for a total of 61 time-bin recurrences, within the optical delay's scanning range, a significant advancement from our prior studies [31]. The measured repetition time of the revivals is 11.03 ps, which corresponds to half the repetition period of the BFC [49], and agrees well with our theoretical modeling. The visibility of the recurrence dips decreases exponentially (see right inset in Figure 1b) due to the Lorentzian lineshape of the BFC frequency bins. In particular, the ≈ 640 ps, from the 1.56 GHz Lorentzian linewidth of the 45.32 GHz cavity, allows us to observe 61 time bins for an approximately 11.03 ps bin spacing. A narrower linewidth would yield even more measurable time bins.

**Frequency-bin revival subspaces:** To further characterize our BFC state, we measure correlations between different signal-idler frequency-bin pairs. In the experiments presented in Figure 2, we use either the 45.32 GHz FSR cavity or the 5.03 GHz FSR cavity. Each pair of frequency bins is selected by a pair of narrow-band filters for the signal and idler photons. As shown in Figure 2a, each filter is composed of a FBG centered at 1316 nm. For the 45.32 GHz cavity measurements in Figures 2b, 2c, the filter has a 300 pm FWHM bandwidth; for the 5.03 GHz cavity results in Figure 2d, the filter has a 100 pm FWHM bandwidth. The coincidence-counting rate is recorded while the filters are set at different combinations of the signal-idler frequency-bin basis. In Figures 2b and 2c, these bins ranged from -2 to +2, with 0 denoting frequency degeneracy. This figure shows that high coincidence-counting rates are obtained only when the filters are set at the corresponding positive-negative frequency bins, according to Eq. (1). We note that, through temporal-state and conjugate-state projection [52], observation of both HOM revivals and frequency-bin correlation verifies the BFC's high-dimensional entanglement. In addition, we investigate the effects of multi-pair emissions on the signal-idler frequency bin cross-talk, as shown in Figure 2c. At ≈ 4 mW pump power, the frequency-bin largest cross-talk increases by 5.4 dB to -6.31 dB compared to the ≈ 2 mW pump power case shown in Figure 2b.

Figure 2d shows the larger number of BFC frequency bins obtained from using the 5.03 GHz FSR cavity and 100 pm bandwidth tunable filters. In this measurement, although the temperature limit of these tunable filters (≈ 100 ºC) bound the number of frequency bins measurable, there are now many more frequency bins compared to the case in Figure 2b. We also note that higher signal-idler frequency-bin cross-talk is observed in the 5.03 GHz cavity due solely to the 100 pm bandwidth of our filter pair, which spans several frequency bins.

We measure and analyze the frequency correlation and time-bin HOM revival subspaces for the BFCs generated with the 5.03 GHz, 15.15 GHz [31] and 45.32 GHz cavities (measurement of HOM revival dips for the 5.03 GHz cavity is shown in Figure S3 of the Supplementary Information). The BFC's number of frequency bins $N_\Omega$ equals $2\pi B_{\mathrm{PM}}/\Delta\Omega$, where $B_{\mathrm{PM}} = 245$ GHz is the FWHM phase-matching bandwidth of the SPDC source, and its number of time bins $N_T$, within an inverse cavity linewidth, equals $\pi/\Delta\omega\Delta T$. Hence their product satisfies $N_T N_\Omega = \pi B_{\mathrm{PM}}/\Delta\omega$ for all three cavities. For the ideal case, in which all the frequency bins are measurable, we found that

$$N_{T,\text{45 GHz}} N_{\Omega,\text{45 GHz}} \approx N_{T,\text{15 GHz}} N_{\Omega,\text{15 GHz}}. \tag{4}$$

where the subscripts label the cavity FSRs, owing to the nearly identical linewidths of the two cavities. In contrast, the time-bin frequency-bin product for the 5.03 GHz cavity should be roughly a factor of three higher, owing to its smaller cavity linewidth. We note that this time-bin and frequency-bin tradeoff for any of our three cavities supports the dense encoding of the time-frequency quantum key distribution.

**High-dimensional hyperentanglement:** Subsequently we modify the BFC output to create polarization entanglement post-selectively, thus generating high-dimensional hyperentanglement for our BFC photon pairs. Figure 3a shows our experimental setup; coupling loss from the fiber-polarizer-fiber for the two output ports is ≈ 1.3 dB and ≈ 1.5 dB respectively. To demonstrate post-selected polarization entanglement, we record the coincidence-count rates while changing the angle of polarizer P$_2$ when polarizer P$_1$ is set at 45º, 90º, 135º, and 180º. As shown in Figure 3b, the measured fringes are well-fitted with sinusoidal curves, with accidentals-subtracted mean visibilities of 97.96 ± 0.41% [calculated using the $(C_{\max} - C_{\min})/(C_{\max} + C_{\min})$ visibility definition for sinusoidal fringes]. The fitted results are used to obtain the correlation functions and the corresponding $S$ parameters. The results are shown in Figure 3c. First, we measure the coincidences at the Clauser-Horne-Shimony-Holt (CHSH) polarizer angles, and then calculate the $S_{\mathrm{CHSH}}$ parameter, which is given by [53]:

$$S_{\mathrm{CHSH}} = |E(\varphi_1, \varphi_2) - E(\varphi_1, \varphi_2') + E(\varphi_1', \varphi_2) + E(\varphi_1', \varphi_2')|, \tag{5}$$

where $E(\varphi_1, \varphi_2)$ is the two-photon correlation function at measurement angles of $\varphi_1$ and $\varphi_2$, respectively. We choose to optimize $S_{\text{CHSH}}$ by using $\varphi_1 = \pi/4$, $\varphi_1' = \pi/2$, $\varphi_2 = 5\pi/8$, $\varphi_2' = 7\pi/8$ [54]. The $S_{\text{CHSH}}$ is found to be $2.686 \pm 0.037$ from those correlation values, which violates the Bell inequality by 18.5 standard deviations. Second, we estimate the maximal achievable $S_{\text{fringe}}$ parameter of $2.771 \pm 0.016$, from the mean visibility of the entanglement-correlation fringes [26]. The $S_{\text{CHSH}}$ and $S_{\text{fringe}}$ parameters are in good agreement, indicating that the hyperentangled state is generated with high quality. Thus, in addition to the 61 time bins, the BFC has high-fidelity post-selected polarization entanglement, further enabling dense quantum information processing.

**Schmidt number analysis and entanglement dimensionality quantification**

**Frequency-binned biphoton frequency comb Schmidt number**

As alluded to earlier, we regard the BFC as providing discrete-variable frequency-binned and time-binned states, depending on how the BFC is measured. Although the frequency-bin correlation measurements and the revivals of our HOM interferometry provide indications of the potential dimensionality of the discrete-variable states, more precise results can come from examining Schmidt-mode decompositions [55, 56], in the frequency- and time-bin basis. In both cases the relevant quantity is that of the Schmidt number $K$ in the specific basis [5], defined as:

$$K = \left(\sum \lambda_n^2\right)^{-1}, \text{where } \sum \lambda_n = 1, \quad (6)$$

with $\{\lambda_n\}$ being the Schmidt-mode eigenvalues.

For the frequency-binned state, these eigenvalues are obtained from the frequency-binned joint spectral amplitude, $\psi(n_s \Delta\Omega, n_I \Delta\Omega)$ which can be obtained by discretizing the BFC's frequency-domain wavefunction $\psi(\omega_S, \omega_I)$, where $\omega_S$ and $\omega_I$ are the signal and idler detunings from frequency degeneracy. We have assumed – based on our high-quality HOM interference and frequency-bin correlation measurements – that our biphoton comb is a pure state. It is challenging experimentally to extract the joint spectral amplitude because such measurements would require reconstruction of the full phase information of the entangled state. Instead, the joint spectral intensity can be more readily measured by performing spectrally-resolved coincidence measurements, as shown in Figure 2. Therefore, we will approximate the joint spectral amplitude by measuring the joint spectral intensity, $|\psi(n_s \Delta\Omega, n_I \Delta\Omega)|^2$, and assuming that:

$$\psi(n_s \Delta\Omega, n_I \Delta\Omega) = \sqrt{|\psi(n_s \Delta\Omega, n_I \Delta\Omega)|^2}. \quad (7)$$

Then, by extracting the Schmidt eigenvalues $\{\lambda_n\}$ from the joint spectral measurements (i.e., the measured correlation matrix, such as Figure 2d), the Schmidt number of the frequency-binned state, $K_\Omega$, can be obtained as shown in Figure 4a. This parameter indicates how many frequency-binned Schmidt modes are active in the biphoton state, and therefore describes its effective dimension [32]. In particular, extracting the Schmidt eigenvalues $\{\lambda_n\}$ from the five resonance-pairs data of Figure 2b for the 45.32 GHz FSR cavity and 2 mW pumping results in a frequency-bin Schmidt number $K_\Omega = 4.31$. For the 45.32 GHz FSR cavity with 4 mW pump power, the data in Figure 2c leads to $K_\Omega = 3.17$, because the increased signal-idler frequency-bin cross-talk drops the purity of each frequency mode, resulting in a smaller Schmidt number. We also note that the increased multi-pair emissions responsible for this additional cross-talk makes the output of the BFC less like a pure biphoton state. For our 15.15 GHz FSR cavity [31], we obtain a $K_\Omega = 8.67$ frequency-bin Schmidt number, where the number of frequency-correlated pairs is limited only by the maximum temperature tuning of our 100 pm FBG filters.

For completeness, we use the frequency-bin data from Figure 2d to extract the frequency-bin Schmidt number for the 5.03 GHz cavity. Using the third panel in Figure 4a, we obtain $K_\Omega = 11.67$ for that cavity. This lower-than-ideal Schmidt number mainly results from the resolution bounds of our 100 pm tunable bandpass filter, but it still demonstrates the scalability of our high-dimensional BFC frequency-binned state. In Figure 4a we compare the extracted frequency-bin Schmidt eigenvalues $\{\lambda_n\}$ for our three cavities.

From our measurements, the diagonal elements of the spectral-correlation matrix (Figure 2d) show the decreasing-envelope behavior of the BFC. Hence we have observed BFC states with Hilbert space dimensionality, $K_\Omega \times K_\Omega$, of at least 16 for the 45.32 GHz cavity, 64 for the 15.15 GHz cavity, and 121 for the 5.03 GHz cavity. Furthermore, the ideal full quantum Hilbert space dimensionality is estimated to be at least 24 (= 4.9 × 4.9) for the 45.32 GHz cavity. For the 15.15 GHz cavity, the ideal full quantum Hilbert space dimensionality is 400 (= 20 × 20); for the 5.03 GHz cavity, the ideal full quantum Hilbert space dimensionality is 1156 (= 34 × 34), where the numbers arise from the detailed theory for the frequency-binned BFC's Schmidt number in Section II of the Supplementary Information.

**Time-binned biphoton frequency comb Schmidt number**

To estimate the Schmidt number for the time-binned BFC state in a manner analogous to what we used for the frequency-binned BFC state would require knowledge of the discretized joint temporal intensity. Because our BFC is generated with continuous-wave pumping, this discretized joint temporal intensity – under the assumption of a pure-state biphoton – is a diagonal matrix with elements $|\Psi(n\Delta T)|^2$, where $n\Delta T$ is the relative delay between the signal and idler photons. We can estimate the discretized joint temporal intensity from our HOM interferometry data, as we now explain. By sampling the BFC state's temporal-domain wavefunction from Eq. (3) at $\tau = n\Delta T$, we get

$$|\Psi(n\Delta T)|^2 = \frac{\exp(-2|n|\Delta\omega\Delta T)}{\sum_{n=-N}^{N}\exp(-2|n|\Delta\omega\Delta T)}, \quad (8)$$

From Section I of the Supplementary Information, we have:

$$V_n = \exp(-|n|\Delta\omega\Delta T)(1+|n|\Delta\omega\Delta T), \quad (9)$$

thus making it possible to find the BFC's joint temporal intensity from HOM interferometry by inverting the one-to-one relation between $|n|\Delta\omega\Delta T$ and $V_n$. Because measuring the discretized joint-temporal amplitude whose singular-value decomposition is the Schmidt decomposition is prohibitively difficult, we assume that it equals the square-root of the discretized joint-temporal intensity, i.e., we use

$$\Psi(n\Delta T) = \sqrt{|\Psi(n\Delta T)|^2}, \quad (10)$$

for the time-binned wavefunction of the BFC, from which it follows that the time-bin Schmidt-mode eigenvalues, $\{\lambda_n\}$, are given by:

$$\lambda_n = \frac{e^{-2\pi|n|/F}}{\sum_{n=-N}^{N} e^{-2\pi|n|/F}} = \frac{\sinh(\pi/F)\exp(-2\pi|n|/F)}{\cosh(\pi/F)-\exp(-(2N+1)\pi/F)}, \text{ for } 0 \leq |n| \leq N, \quad (11)$$

where $F = \Delta\Omega/2\Delta\omega$ is the cavity finesse. The time-binned BFC state's Schmidt number is then found from Eq. (6), with the following theoretical results based on our three cavities' finesses: $K_{T,5\text{ GHz}} \approx 5.16$, $K_{T,15\text{ GHz}} \approx 6.71$, and $K_{T,45\text{ GHz}} \approx 18.30$. By performing a parametric ($|n|\Delta\omega\Delta T$) fit of our experimental data to the $V_n$ expression in Eq. (9), and then applying the result in Eq. (11) and (6), we obtain the experimental values $K_{T,5\text{ GHz}} \approx 5.11$, $K_{T,15\text{ GHz}} \approx 6.56$, and $K_{T,45\text{ GHz}} \approx 18.02$, which agree well with theory.

For the 45.32 GHz FSR fiber cavity, the HOM revivals expand the Hilbert space dimensionality to 324. After combining with post-selected polarization entanglement, the result is a Hilbert space dimensionality of at least 648. Furthermore, we find that the product of the time-binned and frequency-binned Schmidt numbers (when all the frequency bins are measurable, and the measurable HOM time bins run from -340 ps to 340 ps relative delay) is similar for the 45.32 GHz and 15.15 GHz BFCs:

$$K_{T,45\text{ GHz}}K_{\Omega,45\text{ GHz}} \cong K_{T,15\text{ GHz}}K_{\Omega,15\text{ GHz}}. \tag{12}$$

which mimics the BFC time-frequency product relation from Eq. (4) and provides further evidence that our hyperentangled BFC is generated with high fidelity and high Hilbert space dimensionality.

**Conclusions**

In this work, we have demonstrated high-dimensionality hyperentanglement of polarization and time-frequency subspaces using a biphoton frequency comb. We achieve 61 Hong-Ou-Mandel time-bin recurrences, with a maximum visibility of 98.4% (99.9%) before (after) accidental-coincidences subtraction in a stabilized interferometer. In the complementary frequency subspace, the biphoton joint spectral correlation is mapped over 19 frequency-bins, demonstrating the time-frequency duality. We demonstrate the generality of the high-dimensionality over three different cavities, witnessing the qubit joint spectral-temporal correlations across different cavity mode spacings spanning nearly an order-of-magnitude. With the hyperentangled polarization subspace, we achieved a Bell violation up to 18.5 standard deviations and a recurrent maximal achievable correlation-fringe Clauser-Horne-Shimony-Holt *S*-parameter up to 2.771 for the high-dimensional qubit state. Via Schmidt mode decomposition analysis, we describe and quantify the entanglement dimensionality. For example, our 45.32 GHz cavity hyperentangled biphoton frequency comb achieves a time-binned Schmidt number of 18 and Hilbert space dimensionality of at least 648, one of the highest dimensionality in the time-frequency realization to our knowledge.

**Acknowledgements**

The authors acknowledge discussions with Changchen Chen, Andrei Faraon, Charles Ci Wen Lim, Alexander Euk Jin Ling, and discussions on the superconducting single-photon detectors with Vikas Anant. This study is supported by the National Science Foundation under award numbers 1741707 (EFRI ACQUIRE), 1919355, and 1936375 (QII-TAQS).

**Author contributions**

K.-C.C. and X.C performed the measurements and data analysis. M.C.S., A.K., and Y.S.L. contributed to the measurements. K.-C.C., Y.-X.G, and J.H.S. contributed to the theory and numerical modeling sections. T.Z., Z.X., F.N.C.W, and C.W.W. supported and discussed the studies. K.-C.C. and C.W.W. prepared the manuscript. All authors contributed to the discussion and revision of the manuscript.

**Additional information:**

The authors declare no competing financial or non-financial interests. All data needed to evaluate the conclusions in the paper are present in the paper and/or the Supplementary Information. Additional data related to this paper may be requested from the authors.

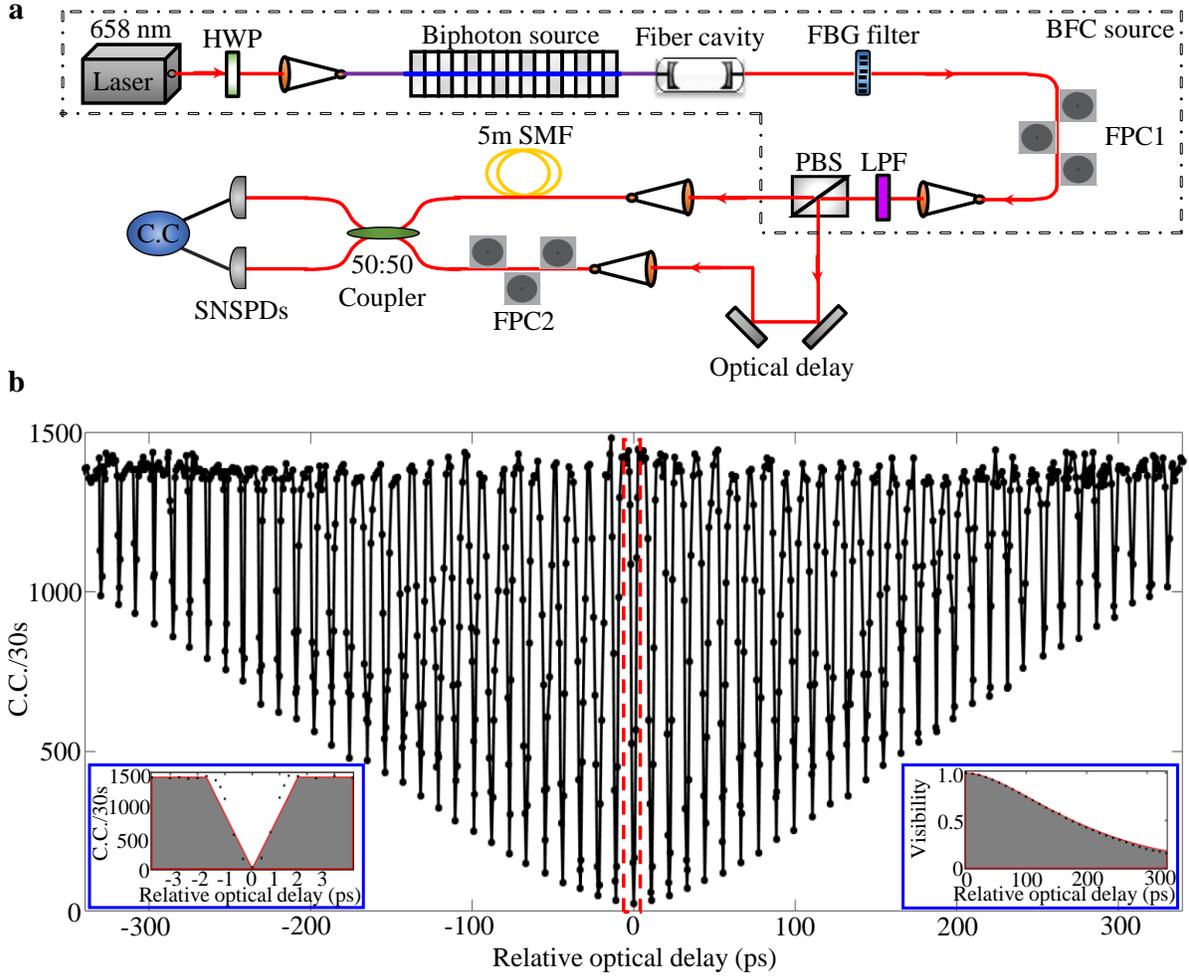

**Figure 1 | High-dimensional BFC generation and HOM quantum-revival observation. (a)** Illustrative experimental configuration. HWP, half-wave plate; FBG, fiber Bragg grating; FPC, fiber polarization controller; LPF, long-pass filter; PBS, polarizing beam splitter; SNSPDs, superconducting nanowire single photon detectors; C.C., coincidence counts. **(b)** Coincidence counts versus relative optical delay, $\Delta T$, between the HOM interferometer's two arms. HOM revivals are observed with up to 61 time-bins. Left inset: zoom-in of the coincidence counts around zero relative delay between the HOM interferometer's two arms. The dip width was fit to 3.86 ± 0.30 ps, which matches well with what is predicted from the 245 GHz phase-matching bandwidth. The central dip's visibility is 98.4%, or 99.9% after subtracting the accidental coincidence counts. Right inset measured time-bin visibility versus the HOM optical delay, compared with theory (red solid line, details in the Supplementary Information).

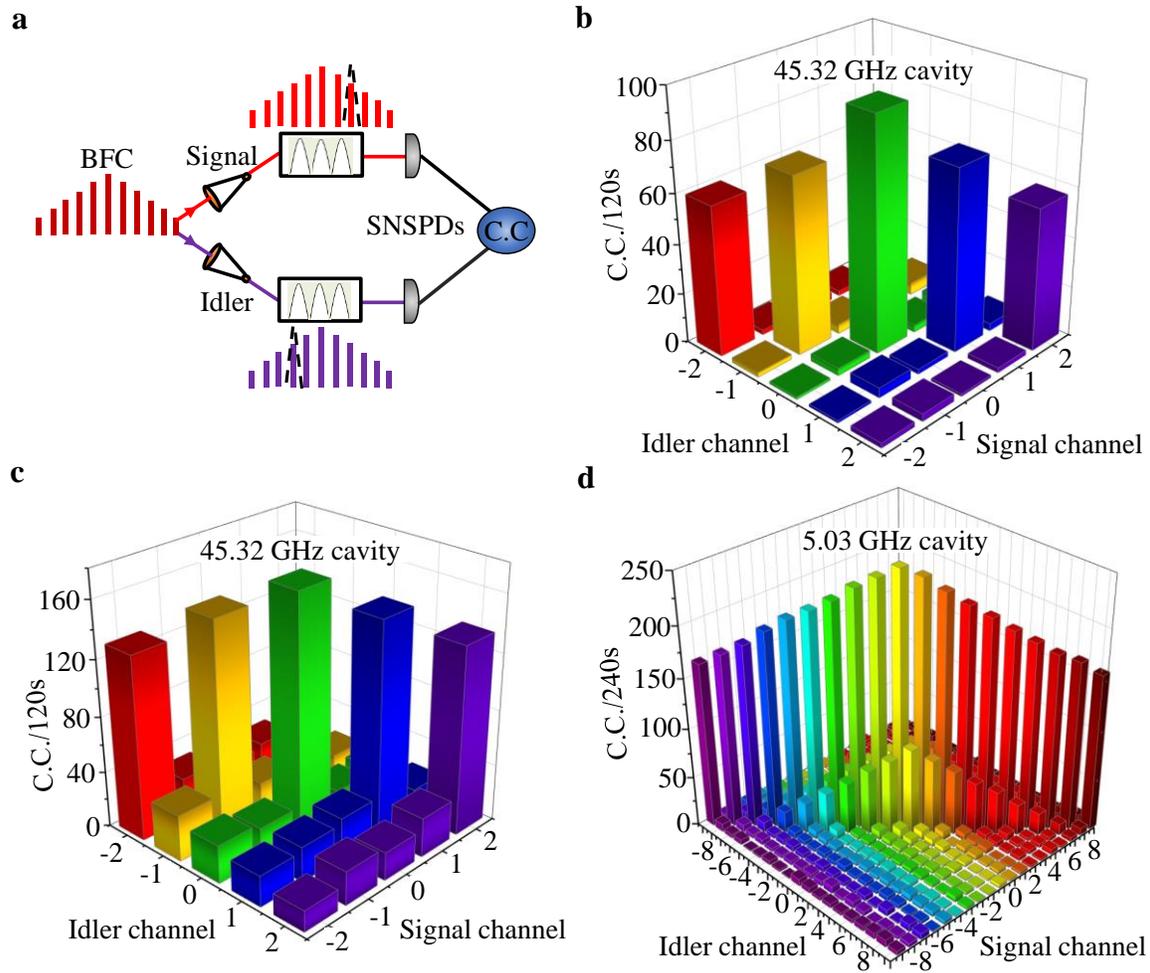

**Figure 2 | Quantum frequency correlations of high-dimensional biphoton frequency combs.**
**(a)** Experimental schematic for the joint spectral intensity measurement for the high-dimensional quantum state. Signal and idler photons are sent to two tunable narrow-band filters for the frequency-bin correlation measurement with coincidence counting. **(b)** Measured frequency correlations of the 45.32 GHz cavity's BFC using filters that had matched full-width-at-half-maximum bandwidths of 300 pm and were manually tuned for scans from the −2 to +2 frequency bins from frequency degeneracy. The SPDC waveguide was pumped at ≈ 2 mW for these measurements, which produced relatively high coincidence counts only along the correlation matrix's diagonal elements. The cross-talk between frequency bins was less than -11.71 dB. **(c)** Measured frequency correlations of the 45.32 GHz cavity's BFC when the SPDC crystal was pumped at ≈ 4 mW, showing increased signal-idler frequency-bin cross-talk of -6.31 dB. **(d)** Higher-dimensional frequency-bin joint spectral intensity measurements for the 5.03 GHz cavity's

BFC. The filters used here had matched full-width-at-half-maximum bandwidths of 100 pm and are temperature tuned for scans from the −9 to +9 frequency bins from frequency degeneracy. The off-diagonal components increase compared to those in Figure 2(a) because the effective bandwidth of the tunable narrow-band filters span multiple frequency bins in this demonstration.

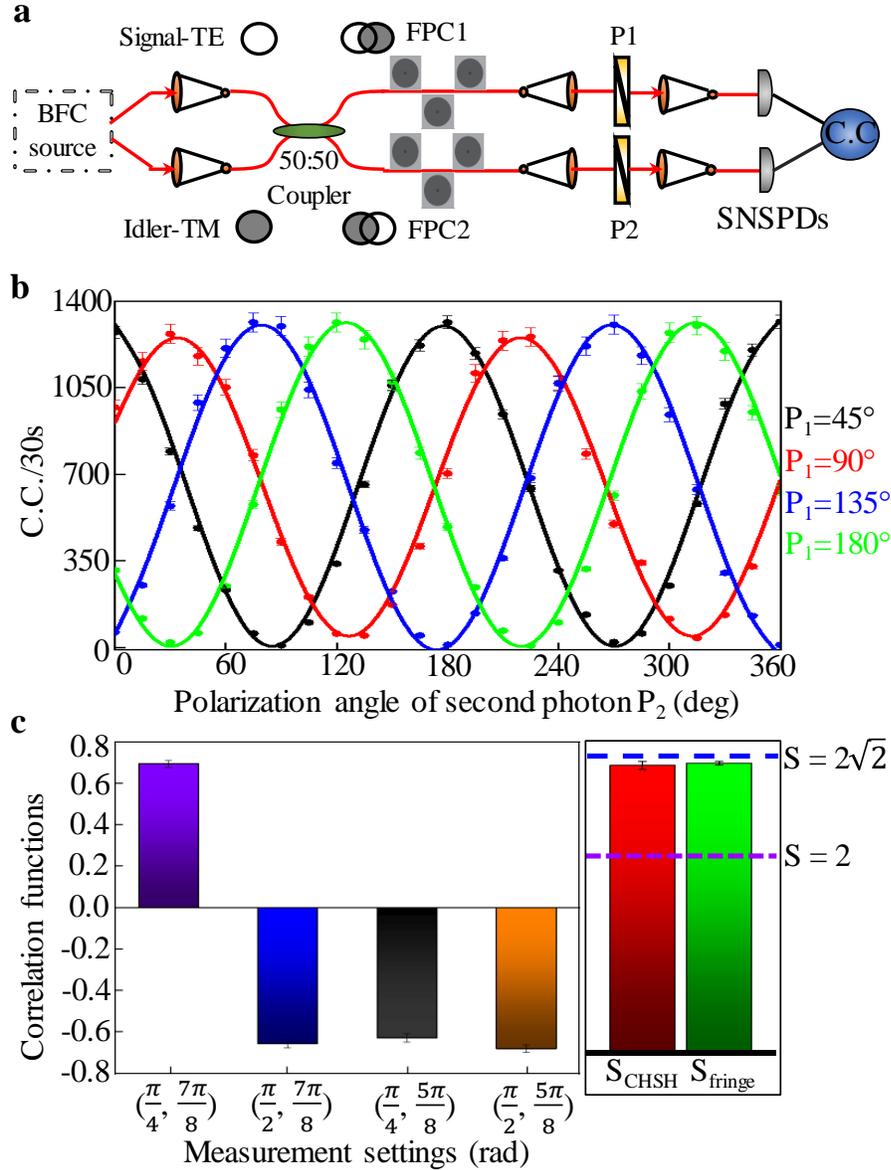

**Figure 3 | Polarization hyperentanglement measurements of high-dimensional biphoton frequency combs.** (**a**) Illustrative experimental scheme. The signal and idler photons are sent to a 50:50 fiber coupler with orthogonal polarizations for post-selected generation of polarization entanglement. P: linear polarizer; C.C.: coincidence counts. (**b**) Polarization entanglement measurements with polarizer P$_1$ fixed at 45º (black curve), 90º (red curve), 135º (blue curve), and 180º (green curve). In both cases, we measured the coincidence-counting rates at the two outputs while changing polarizer P$_2$ from 0º to 360º. In all three cases the measured fringes are well fit with sinusoidal curves, having accidentals-subtracted mean visibilities of 97.96 ± 0.41%. (**c**) Correlation values needed for the Clauser-Horne-Shimony-Holt (CHSH) inequality. The abscissa

label ($\varphi_1, \varphi_2$) denotes the measured polarization bases. The $S_{\text{CHSH}}$ parameter was calculated to be 2.686 ± 0.037 from these correlations, which violates the Bell inequality by 18.5 standard deviations. We obtained the maximal achievable $S_{\text{fringe}}$ parameter to be 2.771 ± 0.016 from the mean visibility of the entanglement correlation fringes. Purple and blue dashed lines denote the classical and quantum boundaries. Error bars represent statistical errors.

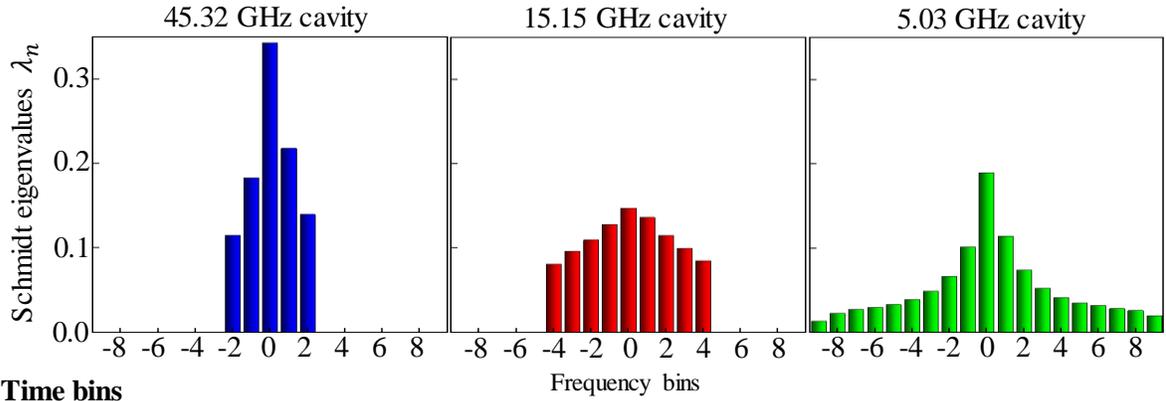
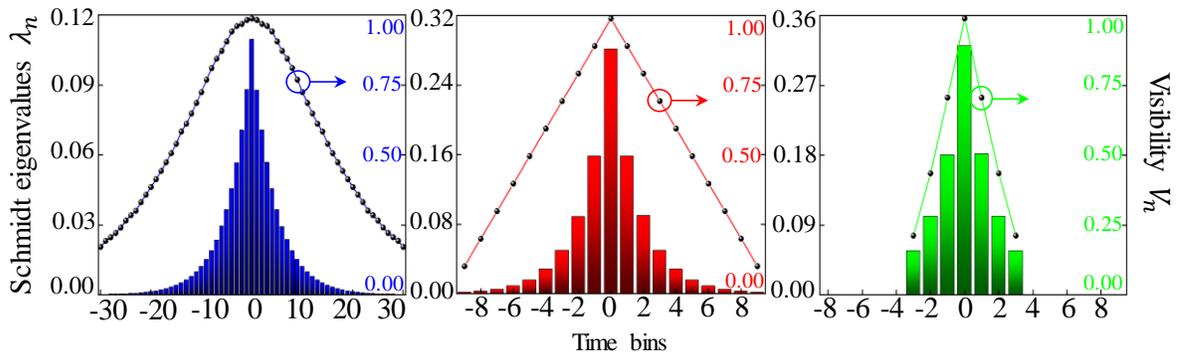

**Figure 4 | Schmidt-mode decompositions for the high-dimensional BFCs.** (**a**) The Schmidt-mode eigenvalues for measured frequency-binned states obtained using our 45.32 GHz, 15.15 GHz, and 5.03 GHz FSR cavities (calculation detailed in the Supplementary Information). (**b**) The Schmidt-mode eigenvalues versus different time bins from the two-photon HOM interferometry. The blue, red, and green bars indicate the discrete time-binned Schmidt number for the 5.03 GHz, 15.15 GHz, and 45.32 GHz cavities, respectively. The central HOM dip is labeled as 0 as a reference.